\documentstyle[12pt,aasms4]{article}
\begin{document}

\title{\bf A Possible Dynamical Effect of a Primordial Magnetic Field}

\author{Jos\'e C.N. de Araujo\altaffilmark{1} and Reuven 
Opher\altaffilmark{2}}

\affil{Instituto Astron\^omico e Geof\'\i sico, Universidade de S\~ao Paulo,\\
Av. Miguel Stefano 4200, S\~ao Paulo, 04301-904 S.P., Brazil}
\altaffiltext{1}{jcarlos@orion.iagusp.usp.br} 
\altaffiltext{2}{  opher@orion.iagusp.usp.br}
\begin{abstract}
The possible existence of a primordial magnetic field in the 
universe has been previously investigated in many articles. 
Studies involving the influence of a magnetic field in the nucleosyntesis 
era, studies considering the effects in the formation of structures 
during the radiation era and the matter era have been considered. 
We here assume the existence of a primordial
magnetic field and study its effect, in particular,
in the formation of voids. The study is twofold:  
to put constraints on the strength of the magnetic field during the
recombination era  and to preview its effects on the formation of voids.
\end{abstract}

\section{Introduction}

The existence of a primordial magnetic field is an old and still open
problem in cosmology (see, e.g., \cite{P93}, for some discussion).
For cosmological scales we may have $<\bf B>=0$, but there might well exist
random magnetic fields at smaller scales (sub-horizon ones) 
such that $<B^2> \ne 0$. 

Magnetic fields have been detected such as in high resolution Faraday rotation
measurements made by Kronberg et al. (1990) of 3C191. The magnetic field 
strength in this z=1.945 system was found to be $\sim 0.4 - 4\mu G$.

Vallee (1990) using a sample of 309 galaxies and quasars obtained
an upper limit of $10^{-9} G \times (0.01/\Omega_{IGM})$ 
(where $\Omega_{IGM}$ is the ratio between the ionized gas density
in the IGM and the critical density, and h is the Hubble constant in units
of 100 km s$^{-1}$ Mpc$^{-1}$) on the strength of the cosmological 
magnetic field which is coherent on horizon scales. 
This limit is reduced to $3\times 10^{-9} G\times(0.01/\Omega_{IGM})$ 
if the cosmic field is coherent only on scales of 10Mpc (see, e.g.,
\cite{KR94}). Magnetic fields of $\sim 10^{-6}$ G is observed to be present 
in spiral galaxies (see, e.g., \cite{KR94}), and also in 
Lyman-$\alpha$ clouds at high redshift, $z\sim 3$ (see, e.g., \cite{P93}).
It is not clear, however, how such fields were produced.

It is worth mentioning that studies of cluster Faraday rotations have been
performed by some authors (see, e.g., \cite{G93}) in trying to look for
primordial magnetic fields.

For galaxies, for example, there is no consensus concerning the origin
of the magnetic field; some authors argue that dynamos could account 
for the growth of putative seed fields, but the magnetic field could well
be primordial. The dynamo is a possible processes to amplify seed magnetic
fields (see, e.g., \cite{V72a}, b among others).

Some authors argued that a magnetic field could provide the primordial
spectrum of density perturbations which would then be gravitationally
amplified (see, e.g., \cite{B96}). In fact, if we consider 
that the magnetic field can be the source for the formation of the structures 
of the universe, we have the same problem that we have
when we try to explain the production of density perturbations, i.e.,
the problem concerning the origin of such a seed magnetic field.

An important question concerning magnetogenesis has to do
with its production epoch; some authors argue that it could 
occur in the inflationary era (see, e.g., \cite{R92}). 
Ratra (1992) showed that fields produced from the inflationary era
could have a strength of $B \sim 10^{-9} G$ at the present time.

Earlier papers by, for example, Harrison (1973) suggested that primordial 
turbulence during the radiation era could produce seed fields, that could be
stocastically amplified, producing at the recombination era $B \sim 1 G$. 
Quashnock et al. (1989) argued that magnetic fields could
be created at the cosmological QCD phase transition. They obtained that
fields corresponding to $B \sim 10^{-17} G$  
by the time of the recombination era could be produced. 

Zweibel (1988) considered a scenario in which density fluctuations, combined
with tidal torques arising between mass condensations, causes regeneration of
a seed field.

The detection of primordial magnetic fields, as argued by some
authors (see \cite{KO96}), is possible.
Kosowsky and Loeb (1996) argued that
primordial fields, that could be present at the last scattering surface,
could induce a measurable Faraday rotation in the polarization of the 
cosmic background radiation. They argued that statistical detection 
of magnetic fields, that would correspond to present day magnetic fields
as small as $10^{-10}$ G, would be possible.

Void regions in the distribution of galaxies are widely accepted
as part of the structure of the universe. The observations show
that  void regions of diameter of up to $\sim 50 h^{-1}$ Mpc are present in
the universe. In a recent study by El-Ad et al. (1996),
studying the new SSRS2 redshift survey (\cite{DA94}),
voids of diameter of $37 \pm 8 h^{-1}$ Mpc were found.

Concerning the formation of voids, three different mechanisms  
have been considered: 1) they may be formed during the formation of the
structures of the universe, in particular, during the clustering processes; 
2) through negative density perturbations; or 3) by explosions 
of pre-galactic objects or quasars.

It is worth mentioning that it is not completely clear how the big 
voids, that range from 10-100 Mpc, have been formed, due to 
the fact that the above three possibilities have problems to account for
very big voids.

In the present work we study a possible dynamical effect that
magnetic fields of a given strength and topology can produce
in the formation of structures, in particular, in the formation 
of void regions. It is not our aim to study MHD in an expanding universe,
some important studies have recently dealt with such an issue (see, e.g.,
\cite{H90})

We are not concerned here with the production of a primordial 
magnetic field. We consider that it could be produced primordially. We 
use strengths that are consistent with the constraints 
imposed by studies concerning the primordial nucleosynthesis and with the 
fields observed in the intergalactic medium and in galaxies.

Studies concerning the effect of a magnetic field
on cosmological nucleosynthesis indicate that fields 
of 0.1 - 1 G at the recombination era are allowed
(see, e.g., \cite{C94,G96}). Considering that we would have at the present 
epoch fields of cosmological origin of $B \sim 10^{-8}-10^{-9}$ G,
we would have at the recombination era  $B \sim 0.001 - 0.1 G$.

We study magnetic fields of strength  $B=0.001 - 1 G$ starting our 
calculations at the recombination
era. We take into account a series of physical
processes, described in detail in the next section, present during and
after the recombination era, as well as the expansion of the universe.
In the present study we consider a baryonic universe, therefore there
is no dark matter (non-baryonic) in our model.

In section 2 we present the model studied, in which we include the basic
equations. In section 3 we present and discuss our results, and finally 
in section 4 we present the main conclusions of our study.

\section{Basic Equations}

Our aim is to study the effects that primordial magnetic fields
could cause in the formation of the structures of the universe. In
particular, our concern in the present work is to study whether primordial 
magnetic fields of a given strength and topology could form void regions. 
On the other hand, our study imposes constraints
on the possible strength of primordial magnetic fields.

The dynamical effects of magnetic fields depend strongly on
their particular topology. In a given cloud the magnetic field
can shrink or expand the cloud.

We choose here a topology that produces an outward pressure; in this
way, the cloud, due to the magnetic thrust, expands.

When magnetic fields are present we use magneto-hydrodynamics to
follow their effects. We consider ideal magneto-hydrodynamics
where the conductivity is infinite, with the magnetic field
frozen into the matter, in such a way that the magnetic flux is conserved
and the changes of the magnetic field are given by:

$$ {\partial\vec B \over \partial t} = \nabla\times(\vec v \times \vec B) $$

\noindent where $\vec v$ is the velocity field of the region where $\vec B$
is present.

We assume that the magnetic field is present
in a finite region of space and within this region we define 
a coordinate system in such a way that the field is in the Z direction.
The topology is chosen in such a way that the magnetic force acts in
the R (radial cylindrical coordinate) direction, in particular, in the 
outward direction. Such a field can be written as: 

\begin{equation}
\vec B=B_{rec}\biggl({ \bar R_{rec}\over \bar R_{\quad}}\biggr)^2 
\sqrt{1-\Big( {R \over \bar R}\Big)^2} \hat Z
\end{equation}

\noindent where: $B_{rec}$ is the magnetic field along the Z axis at the
recombination era, $\bar R_{rec}$ the radius of a
spherical region that encompasses the field region at the recombination era,
$\bar R$ the equatorial radius of the region, and R the radial cylindrical 
coordinate. Note also that this magnetic field goes to zero at the finite
radius $\bar R$ as $\sqrt{1-(R/\bar R)^2}$. Such a field has obviously an 
axial symmetry, but we are going 
to consider that it is present in a region that we 
consider to be initially spherical. This is in fact a 
rough approximation but it gives information on how such a field can 
influence that region. To follow in detail an actual situation
in which a magnetic field is present, we need to consider a series 
of effects very difficult to deal with, even considering very 
complex models that involve the three spatial coordinates and time.

The magnetic field given by Eq. 1 satisfy the equation for flux conservation
if $\vec v \propto \vec R$. A velocity field linear in $\vec R$ (and also
in $\vec Z$) is consistent with a cloud contracting or expanding with 
uniform density. This very fact has been the main reason for the particular
topology that we have chosen. As a result, as we will see below, the
set of differential equations to be solved depends only on time.

Mestel (1965), for example, in his study on the problem of star formation in
the presence of a magnetic field, also adopted the same topology that we have
adopted here. He did so assuming the same argumentation that we have used, 
i.e., to maintain the density uniform where the field is present.

Still concerning Eq. 1, it represents a generic simple magnetic topology of
the simplest current configuration of a magnetic fluctuation discussed in the
literature - a current loop. A magnetic field configuration is associated
with a current configuration from Maxwell's equations. A current loop tends
to expand. The opposite sides of the loop repel each other by the Biot-Savart
law. Instead of treating the expansion of the current loop it is easier to
treat the expansion of the magnetic field that it creates due to the magnetic
pressure acting on the external medium. Thus the expansion due to the magnetic
pressure of the magnetic topology of Eq. 1 describes the expansion of a 
simple current configuration such as the current loop which expands due to
the Biot-Savart law. A simple current loop is very stable. The universe is
highly conductive and current loops greater than $\sim 10^{13} cm$, by the
time of the recombination era, did not dissipate on a Hubble time (see, e.g.,
\cite{C94}).

To follow the region in which the field is present we must
solve the magneto-hydrodynamic equations, namely:

\begin{equation}
{\partial\rho \over \partial t} + \nabla . \rho \vec v = 0 
\end{equation}

\noindent the continuity equation;

\begin{equation}
{d\vec v \over dt}+ \nabla\phi +{1\over\rho}\nabla(P+P_m) +
{\sigma b T^4_r \over m_pc}x(\vec v - H\vec r) = 0
\end{equation}

\noindent the equation of motion, where the last term is the
photon-drag due to the cosmic background radiation;

\begin{equation}
\nabla^2\phi = 4\pi G\rho
\end{equation}

\noindent the Poisson equation for the gravitational potential;

\begin{equation}
P=N\rho kT(1+x)
\end{equation}

\noindent the equation of state;

\begin{equation}
P_m={ B^2 \over 8\pi}
\end{equation}

\noindent the magnetic pressure; and

\begin{equation}
{du\over dt}= -L + {P \over \rho^2}{d\rho \over dt}
\end{equation}

\noindent the energy equation.

\noindent 
In the above equations: $\rho$ is the matter density, $\vec v$ the 
velocity, $\phi$ the gravitational potential, $P$ the matter pressure,
$P_m$ the magnetic pressure, $ H\vec r$ the Hubble flow, 
$H=\dot a /a$ with $a$ and $\dot a$ being the scale factor and 
its time derivative, respectively, $T_r$ the background radiation 
temperature, $T$ the matter temperature, $x$ the degree of ionization, 
$u$ the energy density, $L$ the cooling function, $\sigma$ the 
Thomson cross section, $b= 4/c$ times the Stefan Boltzmann constant, 
$m_p$ the proton mass, $c$ the velocity of light, $G$ the 
universal gravitational constant, $N$ the Avogrado's number, and 
finally $k$ the Boltzmann constant.

For the degree of ionization we follow the article by Peebles (1968)
in which the processes of recombination are taken into account in 
detail. 

In the cooling function a series of processes have been taken
into account: photon cooling (heating), recombination,
Lyman-$\alpha$, and $H_2$ molecules. In the present work we obtained
that the most relevant mechanism is photon cooling (heating).

It is worth stressing that it is not necessary to consider the putative losses
via the presence of magnetic fields in the present study. It could be 
important in principle, but it is the photon-cooling (heating) that dominates
the cooling-heating processes throughout the recombination era. These very
processes maintain the matter temperature ($T_m$) close to the temperature
of the background radiation ($T_r$) until the decoupling time. Yet,
the thermal pressure is not important in the formation of the void region,
the main mechanism that influences the formation of void region is the
photon-drag (besides the magnetic thrust) that tries to inhibit any relative
motion between the baryonic matter and the background radiation. The 
photon-drag depends on the radiation temperature to the fourth power and 
linearly on the degree of ionization. The degree of ionization depends on
(during the recombination era) the photoionization rate (that depends on
$T_r$) and on the recombination rate (that depends on $T_m$ ($\sim T_r$)).

After the decoupling time, other processes could be important, in particular
in what concerns the thermal evolution of the void region. However, the size
of such a void region does not depend on these other processes, since the
thermal pressure is not important, and by this time neither the magnetic
pressure is important.

To resolve the above set of equations we integrate it only
in time. This is so due to the following reasons.        
First, the magnetic force, the photon-drag and the gravitational
force depend on R and the Z coordinates linearly inside the
region where the field is present. This implies that the velocity 
depends linearly on the above mentioned coordinates
multiplied by a function of time.
Second, due to the fact that the linear profile in the velocity
is maintained throughout the region of interest the density profile
is, as a consequence, uniform. The density in the region of interest
as a consequence depends only on time. Similar models, but without taking 
into account the presence of magnetic fields, were developed by
de Araujo and Opher (1989, 1990, 1991, 1993 and 1994) to study the
formation of Population III objects, galaxies and voids. In fact,
the difference between the present work and our latter studies
is the inclusion of the magnetic term in the equation of motion.

From the above we see why the particular topology has been chosen. The aim
is to have a magnetic pressure gradient producing an outward force that
would maintain the density profile uniform in the region where the field
is present. Any other topology produces a magnetic force that destroys
the ``top-hat'' profile. To follow the evolution of the region where the
field is present would then require the integration of the hydrodynamic
equations not only on time but also on the R and Z coordinates.

To proceed, we resolve the set of equations writing the density in the form

\begin{equation}
\rho (t)=\bar\rho (t) +\delta\rho (t)=\bar\rho (1+\delta)
\end{equation}

\noindent where $\bar\rho =\rho_o(a_o/a)^3$, $\rho_o$ is the present density,
a the scale factor, and $\delta\equiv\delta\rho/\bar\rho$. Note that we are
assuming a ``top hat'' density profile. Note also that there is no incoming
or outgoing of material, since we take a fixed value for M.

We use a linear velocity profile for the cloud (which is consistent with
constant spatial profile),

\begin{equation}
\vec v=\vec v_H+v_{1R}(t){\vec R \over \bar R}+v_{1Z}(t){\vec Z \over \bar Z}
\end{equation}

\noindent where  $\vec v_H = \vec r (\dot a/ a)$  (the Hubble flow).

The resulting set of equations for $\delta (t)$, $v_{1R}(t)$ (and $v_{1Z}(t)$)
and $T_m$ are very similar to Eqs. 15, 16 and 19 of de Araujo and Opher 
(1989). They have been only adapted to the cylindrical coordinates used here,
and in the equation for $v_{1R}$ we have included the magnetic pressure.
The equation for the degree of ionization is taken from Peebles (1968).

The system of equations that we have derived are first order differential
equations that we integrate on time using a Runge-Kutta IMSL routine.

\section{Calculations and Discussion}

We have performed a series of calculations for a baryonic universe 
with $\Omega_0=0.1$ (where $\Omega_0$ is the baryonic density parameter) 
and h = 1.0 (the Hubble constant in units of 100 km s$^{-1}$ Mpc$^{-1}$). 
We begin the calculations at the beginning of the recombination era, starting from the time when the
temperature of the radiation is $T_r=4000K$ (or at the redshift z=1480). 

We consider that the magnetic field at the beginning of our calculations 
ranges from $B=10^{-3} - 1$G for different scales. 
If such a field decreased from the recombination era until today
only as a function of the scale factor ($B \propto a^{-2}$),
we would have today fields with strengths ranging from 
$4.6\times 10^{-7}-4.6\times 10^{-10}$ G. These fields however thrust
the matter around it, and in consequence the region where the field is 
present expands at a rate greater than the expansion rate of the
universe. Due to the fact that the field is frozen into the matter it 
varies as $B \propto \bar R^{-2}$  (instead of $B \propto a^{-2}$). Our
calculations show that we have today fields $\sim 10^{-9}-10^{-11}$ G.

The scales are defined initially by the diameter of the clouds encompassing 
masses in the range $10^8-10^{14} M_\odot$, which give scales in the 
range $\sim 0.1 - 12kpc$, that are smaller than the horizon at the beginning 
of the recombination era.

In general, the effect of the magnetic field with the topology chosen
here is to thrust the matter, and this effect is very strong during
the recombination era. After the recombination era the dynamical effect of 
such a field decreases. 

In Table 1 we present our main calculations; the first column
names the models, the second the magnetic field at the beginning of
the recombination era, the third the present magnetic field, the 
fourth the initial radius of the region where the field is present,
the fifth the mass contained in the region where the field is present,
the sixth and the seventh the present equatorial and polar radii of the 
void, respectively, and finally the eighth column gives
the present density contrast. The density contrast is defined as:

\begin{equation}
\delta\equiv {(\rho_{void} -\bar \rho) \over \bar\rho}
\end{equation}

\noindent where $\rho_{void}$ is the void density and
$\bar \rho$ is the density of the universe.

Depending on the initial dimensions of the region, a field 
of 1 G can produce a void region of up to 200 Mpc. 
As stressed above, the present work helps to impose
constraints on the strength of the primordial magnetic field.
This result thus suggests that fields of 1 G cannot be 
present in regions of $\sim 6 kpc$ at the time of the recombination era.

$B_{rec}$ = 0.1 G could produce voids of semi-major 
axis ranging from  $\sim  5-34$ Mpc, which could either account for 
the big voids observed, or, again, impose limits on the strength of 
magnetic fields present in such dimensions.

Our calculations suggest that $B_{rec}$ = 0.01 G  cannot account
for voids of dimensions larger than $\sim 5$ Mpc. If, for example, a field of
0.01 G is present in regions of $ \sim 6$ kpc at the beginning of the
recombination era it produces $\delta_p=-0.15$, which is not enough
to be a void.

In table 1 we also show that fields of $B_{rec}$ = 1 mG cannot produce 
voids larger than $ \sim 1$ Mpc, assuming that to have voids it
is necessary to have $\delta_p < -0.5$. 

From our calculations we can conclude that if primordial magnetic
fields are present at the recombination era, it can produce a
dynamical  effect, in particular, in the formation of void regions,
if its strength is larger than $\sim$ 1 mG.

To illustrate how the magnetic fields, with the particular
strengths and topology investigated, could form void regions, we show
in figure 1 the density contrast as a function of redshift (z),
in particular for the model D3. It is shown that after
a negligible decrease in $\delta_p$, the magnetic pressure
efficiently thrusts the matter and produces a void. In
the beginning, the increase of the void region is difficult
to occur due to the background radiation that tends to inhibit any motion
of matter in the beginning of (as well as before) the recombination era. 

In figure 2 we show the change of the semi-major and semi-minor axes 
as z decreases. In figure 3 we illustrate how
the magnetic field evolves during the formation of the void.

\section{Conclusions}

In the present work we study a possible dynamical effect that could
be produced if a primordial magnetic field is present at the time
of the recombination era with the strength and topology assumed here.
The study is two-fold: 1) to show that a magnetic field can
be important in the formation of structures in the universe, in particular,
in the formation of void regions; and 2) to impose constraints on its
presence at the time of the recombination era.

For a baryonic universe with $\Omega _0=0.1$ and h=1.0, we
show that void regions of diameter of $\sim 60$ Mpc could
be formed if $B_{rec}$=0.1 G. We also obtain that magnetic fields
of $B_{rec} < 1$ mG would not produce any significant dynamical effect.

Obviously the particular effects obtained here depend strongly on 
the particular topology that we have chosen. We here have a 
repulsive magnetic force $\propto R$. Other topologies could produce
repulsive forces  $\propto R^\beta$. If, for example, $\beta > 1$ the magnetic force
would produce stronger effects.

In the present study, as already mentioned, we have not taken into 
account the presence of non-baryonic dark matter. An interesting 
study would be to take into account the presence of non-baryonic dark matter 
in order to see its influence on the conclusions present here. 
It is our aim in the future to address such a study.

\acknowledgments

The authors would like to thank the Brazilian agency CNPq for support.
JCNA would like to thank Oswaldo D. Miranda for helpful discussions. Finally,
we would like to thank the referee for helpful comments.

%
%
%
%

\clearpage
\begin{deluxetable}{cccccccc}
\tablecaption{For different values of $B_{rec}$ and dimensions, we
show the present values of: the~magnetic field ($B_0$), the polar ($Z_p$) and 
the equatorial ($R_p$) radii, and the~density~contrast~($\delta_p$) for
a baryonic universe with $\Omega_0=0.1$ and h=1.0.}
\tablewidth{0pt}
\tablehead{
\colhead{Model} & \colhead{$B_{rec}$} & \colhead{$B_0$} &
\colhead{$\bar R_{rec}$} & \colhead{$M/M_\odot$} & 
\colhead{$R_p$} & \colhead{$Z_p$} & \colhead{$-\delta_p$} \\
\colhead{ } & \colhead{$(Gauss)$} & \colhead{$(Gauss)$} &
\colhead{$(kpc)$} & \colhead{ } & 
\colhead{$(Mpc)$} & \colhead{$(Mpc)$} & \colhead{}
 }
\startdata

$A1$ &$0.001 $& $4\times 10^{-11}$& $0.064 $& $10^{8\,}$& $0.34$& $0.30$ & $0.97503$ \nl
$A2$ &$0.01  $& $1\times 10^{-11}$& $0.064 $& $10^{8\,}$& $2.04$& $1.63$ & $0.99987$ \nl
$A3$ &$0.1   $& $2\times 10^{-11}$& $0.064 $& $10^{8\,}$& $4.99$& $3.14$ & $0.99999$ \nl 
\ \nl
$B1$ &$0.001 $& $3\times 10^{-10}$& $0.30  $& $10^{10} $& $0.55$& $0.54$ & $0.47138$ \nl
$B2$ &$0.01  $& $1\times 10^{-10}$& $0.30  $& $10^{10} $& $3.06$& $2.66$ & $0.99653$ \nl
$B3$ &$0.1   $& $4\times 10^{-11}$& $0.30  $& $10^{10} $& $14.4$& $11.1$ & $0.99996$ \nl 
$B4$ &$1.0   $& $9\times 10^{-11}$& $0.30  $& $10^{10} $& $31.3$& $18.8$ & $1.00000$ \nl 
\ \nl
$C1$ &$0.001 $& $4\times 10^{-10} $& $1.39 $& $10^{12} $& $2.08$& $2.08$ & $0.35726$ \nl
$C2$ &$0.01  $& $1\times 10^{-9\,}$& $1.39 $& $10^{12} $& $3.89$& $3.65$ & $0.84345$ \nl
$C3$ &$0.1   $& $3\times 10^{-10} $& $1.39 $& $10^{12} $& $26.2$& $22.0$ & $0.99943$ \nl 
$C4$ &$1.0   $& $2\times 10^{-10} $& $1.39 $& $10^{12} $& $89.3$& $64.9$ & $0.99998$ \nl 
\ \nl
$D1$ &$0.001 $& $5\times 10^{-10} $& $6.43 $& $10^{14} $& $9.53$& $9.53$ & $0.00171$ \nl
$D2$ &$0.01  $& $4\times 10^{-9\,}$& $6.43 $& $10^{14} $& $10.1$& $10.0$ & $0.14951$ \nl
$D3$ &$0.1   $& $4\times 10^{-9\,}$& $6.43 $& $10^{14} $& $33.8$& $30.5$ & $0.97523$ \nl 
$D4$ &$1.0   $& $1\times 10^{-9\,}$& $6.43 $& $10^{14} $& $206 $& $166 $ & $0.99988$ \nl 

\enddata
\end{deluxetable}

\clearpage
\begin{figure}
\epsscale{0.84}
\plotone{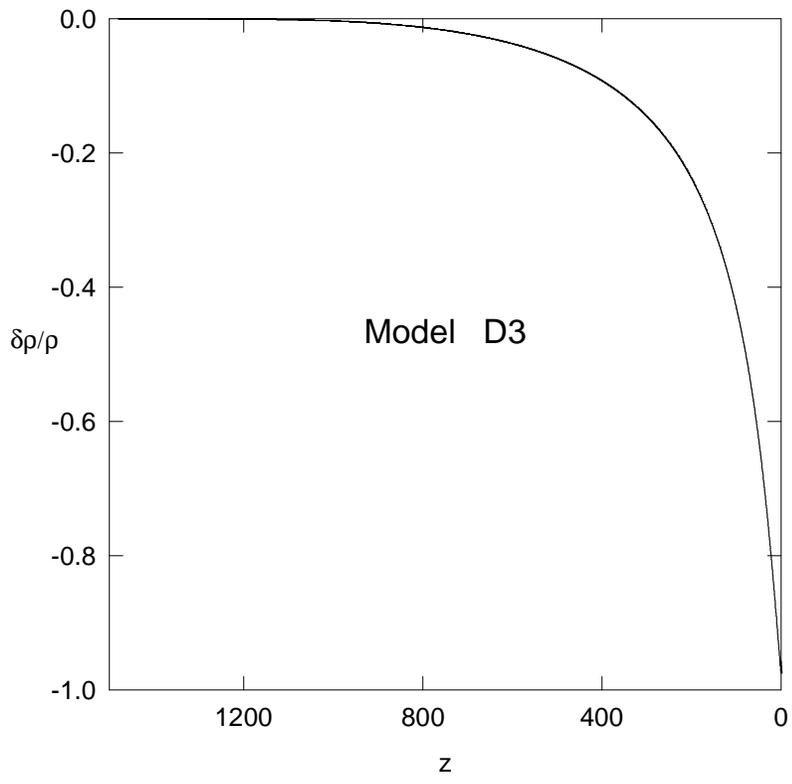}
\caption{$\delta\rho/\rho$ as a function of redshift.}
\end{figure}

\clearpage
\begin{figure}
\plotone{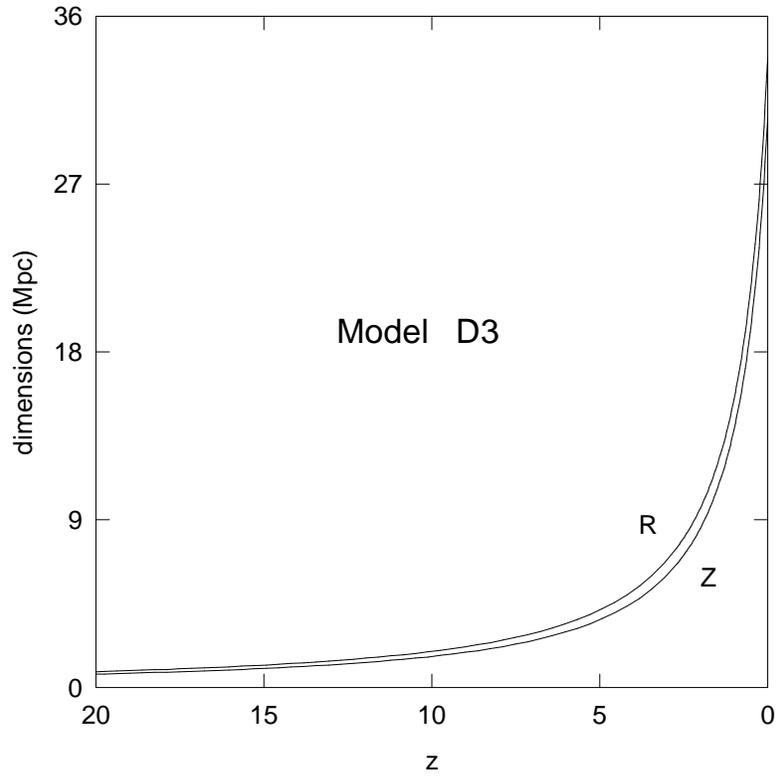}
\caption{Equatorial (R) and Polar (Z) radii as a function of redshift.}
\end{figure}

\clearpage
\begin{figure}
\plotone{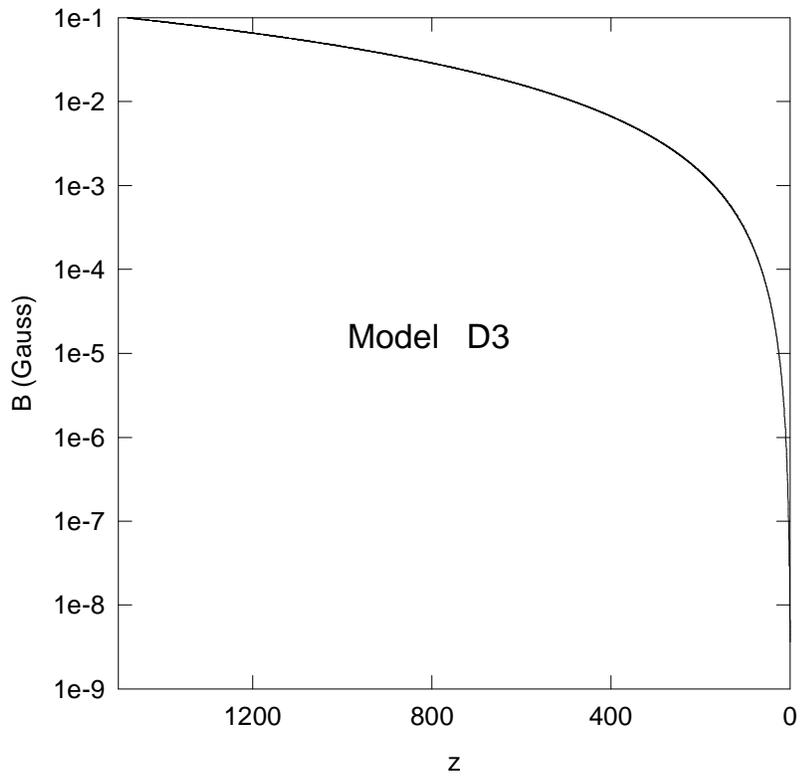}
\caption{The evolution of the magnetic field as a function of redshift.}
\end{figure}

\end{document}